\documentclass[12pt]{article}
\usepackage{amsmath,amssymb,graphicx,mathrsfs,hyperref,slashed,color}
\newcommand{\be}{\begin{equation}}
\newcommand{\ee}{\end{equation}}
\newcommand{\bea}{\begin{eqnarray}}
\newcommand{\eea}{\end{eqnarray}}

\newcommand{\BH}{{\mbox{\tiny BH}}}
\newcommand{\mrm}{\mathrm}

\def\({\left(} \def\){\right)}

\renewcommand{\baselinestretch}{1.25}
\begin{document}
\title{\vspace{-1.8in}
{When black holes collide: Probing the interior composition by the spectrum of ringdown modes and emitted gravitational waves}}
\author{\large Ram Brustein${}^{(1)}$,  A.J.M. Medved${}^{(2,3)}$, K. Yagi${}^{(4)}$
\\
\vspace{-.5in} \hspace{-1.5in} \vbox{
 \begin{flushleft}
  $^{\textrm{\normalsize
(1)\ Department of Physics, Ben-Gurion University,
    Beer-Sheva 84105, Israel}}$
$^{\textrm{\normalsize (2)\ Department of Physics \& Electronics, Rhodes University,
  Grahamstown 6140, South Africa}}$
$^{\textrm{\normalsize (3)\ National Institute for Theoretical Physics (NITheP), Western Cape 7602,
South Africa}}$
$^{\textrm{\normalsize (4)\ Department of Physics, Princeton University, Princeton, New Jersey 08544, USA}}$
\\ \small \hspace{1.07in}
    ramyb@bgu.ac.il,\  j.medved@ru.ac.za,\ kyagi@princeton.edu
\end{flushleft}
}}
\date{}
\maketitle
\begin{abstract}

The merger of colliding black holes (BHs) should lead to the  production of ringdown or quasinormal modes (QNMs), which may very well be sensitive to the state of the interior. We put this idea to the test with a recent proposal that the interior of a BH consists of a bound state of highly excited, long, closed, interacting strings; figuratively, a collapsed polymer. We show, using scalar perturbations for simplicity, that such BHs do indeed have a distinct signature in their QNM spectrum: A new class of modes whose frequencies are  parametrically lower than the lowest-frequency  mode of a classical BH  and whose damping times are parametrically longer. The reason for the appearance of the new modes is that our model contains another scale, the string length, which is parametrically larger than the Planck length. This distinction between the collapsed-polymer model and general-relativistic BHs could be made with gravitational-wave observations and offers a means for potentially measuring the strength of the coupling in string  theory. For example, GW150914 already allows us to probe the strength of the string coupling near the regime which is predicted by the unification of the gravitational  and  gauge-theory couplings. We also derive bounds on the amplitude of the collapsed-polymer QNMs that can be placed by current and future gravitational-wave observations.

\end{abstract}
\newpage
\renewcommand{\baselinestretch}{1.5}\normalsize

\tableofcontents

\section{Introduction and Summary of Results}

The narrative of classical general relativity (GR) is that the interior of a  Schwarzschild black hole (BH) is a region of  empty space surrounding a classically singular center. Recently, this picture was understood to be in contradiction with the laws of quantum mechanics and  thus revealed as  misleading. The  modern alternative  scenario is that the interior does not exist and spacetime comes to an abrupt end at the BH horizon ---  either physically as in the fuzzball model of BHs \cite{Mathur1,Mathur2,Mathur3,Mathur4} (also  \cite{otherfuzzball} and, more recently, \cite{hooft}) or effectively as a ``firewall'' of high-energy particles surrounding the horizon \cite{AMPS} (also \cite{Sunny,Braun,MP}). This scenario suggests that at least the near-horizon state (and perhaps the whole interior)
has to deviate substantially from the vacuum; a situation that differs greatly from the expectations of GR. The degree of deviation is still under debate.

Here, we will be adopting a model of a Schwarzschild BH for which the interior is not mostly empty, in stark contrast to the GR case. The BH interior rather contains a particular state of matter: a non-classical,  bound state of  long, closed, highly excited, interacting  strings; in essence, a collapsed polymer \cite{strungout}. A more figurative way of describing  the bound state might be as a ``quantum star" consisting of hot fundamental strings in the Hagedorn phase or  simply
as a  ``string ball". A more detailed account of this collapsed-polymer model is provided in an appendix, see Sec.~A.1. The polymer's outer surface acts just like a classical BH horizon in the limit $\;\hbar\to 0\;$; that is, the interior and exterior are causally disconnected in that  enclosed matter had no opportunity to escape from the interior. However, this is only approximately true once quantum effects have been ``turned on'' \cite{emerge}.

We have  argued elsewhere that the low occupation numbers of the Hawking radiation along with the assumption of a unitary theory necessitates a strongly non-classical state  of matter within the  BH interior  \cite{density,inny,noclass}. Given such a state,  a geometric mean-field  description in terms of a  metric and other spacetime fields  is no longer feasible.  But then, faced with  the absence of an effective description of the geometry, what can one actually say about the interior of a Schwarzschild BH and its influence on the exterior? We will eventually  show that the composition of the interior does indeed become relevant in the context of BH mergers.

Some of our results are expected to hold in general for BH-like  objects.
For us, ``BH-like objects'' represents a collective name for exotic spacetimes containing ultra-compact  objects  that can mimic some of the basic properties of a BH as viewed by an external observer but without conforming to the picture from GR  (mostly empty space with a dense, singular core).  These objects include, for example, wormholes, gravastars, firewalls, fuzzballs, graviton condensates, boson stars, neutron stars with a certain equation of state or an anisotropic pressure and, of course, collapsed polymers. Some of the objects, such as boson stars, do not possess an essential property of BHs: a horizon (even an effective one), meaning a  bounded  region of spacetime from which matter cannot escape classically.

But, as far as we are aware, any known form of matter  cannot support such Schwarzschild-sized  objects without collapsing under the influence of gravity \cite{buchdahl,Fujisawa:2015nda}. This is because all known  interactions of standard matter are weaker than gravity under these circumstances. Only highly excited string matter seems to be capable of supporting  compact enough objects with the properties of a BH and yet not collapse any further. This is the impetus for our current focus on the polymer model; nevertheless,  a companion paper \cite{spinny} considers a more general class of (rotating)  BH-like objects.

Our objective is to show how   gravitational waves (GWs) can be used
as a means for distinguishing  the collapsed-polymer model from  classical
BHs and from other BH-like objects.
An observable signal of GWs can be produced from the merger
of two colliding BHs.
Such an  event  proceeds in  three  stages: the inspiral (or pre-merger), merger  and  post-merger (or ringdown) stages. In the last of these, the newly formed
BH will settle down by emitting  ringdown modes ---
also known as quasinormal modes (QNMs) ---  which are physically realized in the form of GWs. Note, however, that our analysis uses scalar perturbations for simplicity.

It would be difficult to use the early part of the inspiral stage to discriminate various BH-like models because its binary components are adequately described by point particles. On the other hand, one could, in principle, use the tidal information
which is encoded in the later part of the inspiral stage to probe BH-like objects~\cite{Wade:2013hoa,Yagi:2016ejg,Mendes:2016vdr,Uchikata:2016qku,Cardoso:2017cfl}. More dominant effects in terms of post-Newtonian order counting for the purpose of probing exotic compact objects include the quadrupole moment \cite{quad} and tidal heating at the horizon \cite{Maselli:2017cmm}. But the merger phase is complicated by its highly non-linear evolution.
Moreover, we currently lack merger simulations of binary BH-like objects
(except for boson stars~\cite{Palenzuela:2007dm,Mundim:2010hi}) that would
enable
us to probe the merger stage  for these exotic spacetimes. Fortunately, the post-merger  stage can provide us with an excellent  opportunity for detecting   QNMs,  thanks to the recent advances in GW  observations and the promise for future detections~\cite{LIGO,Abbott:2016nmj,TheLIGOScientific:2016pea,LIGO1}.
A discussion on QNMs can be found in Sec.~A.2.

\subsection{Previous work on constraining exotic spacetimes from GW150914}

Let us recall here the analysis of the famous merger event GW150914 by the
LIGO and Virgo collaboration \cite{LIGO}, as well as  an associated analysis which  constrains  possible exotic spacetimes \cite{LIGO1,Chirenti:2016hzd,Konoplya:2016pmh,pretorius}.

It is generally fair to say that the constraints, in cases for which they apply, are currently weak. The statistical significance in the detection of the merger comes mostly from the pre-merger and merger phases, whereas  that of the ringdown phase is not so useful. What little is known about the  ringdown phase is, however,  consistent with GR. But  this by itself does not have a strong discriminating power among the predictions of GR and  various  BH-like candidates because, as discussed in Sec.~A.2, a sufficiently compact object should be able
to produce modes that closely resemble  the predominant modes of GR.

Given that the LIGO and Virgo Collaboration did not report the presence of a secondary ringdown mode, Yunes and collaborators \cite{pretorius} have placed interesting bounds on the intrinsic dissipation,  ringdown frequency $f_{RD}$ and damping frequency $f_{damp}$ of applicable BH-like objects. However, the region of small frequency --- our region
of interest --- was not covered by their analysis.

\subsection{Summary of results}

We will show in what follows  that the collapsed-polymer model predicts
a novel class of  low-frequency, long-lived modes.   The frequencies of this class are parametrically lower than the GR scale  $c/R_S$ (the inverse of the ``Schwarzschild time'')  by a factor of the string coupling $g_s$; that is,  $\;\omega_{R} \sim g_s~c/R_S\;$, and  the damping times are longer than $R_S/c$ by a factor of the  square of the string coupling, $\;\tau_{damp}\sim 1/g_s^2~R_S/c\;$. The estimate from the quadrupole formula implies (albeit with less certainty) that the expected strain of the emitted GWs is smaller by a factor of $(g_s^2)^2$ than the strain of the conventional GR modes.

The string coupling is small but, in many string theories and  models, it is not ``very small''. For instance,
in string theory, if one requires the unification of the gravitational  and  gauge-theory couplings, the expectation is $\;g_s^2/4\pi =1/25\;$ or $\;g_s^2 \simeq 1/2\;$ \cite{unif}.  One can just as easily imagine other scenarios in which $\;g_s^2 \sim 1/100\;$ or even smaller, but it is not related to any of the extremely small parameters of the BH such as $1/S_{BH}$ ($S_{BH}$ is the BH entropy). Therefore, the value of $g_s^2$ could easily fall somewhere between $1/2$ and $1/100$. Thus,  there is the tantalizing possibility that a mode is detected whose frequency is lower than those of the GR modes, and whose delay in emission time is long enough to be definitive but still short enough to be observationally relevant to  future experiments. In this way, there is a characteristic signature for the polymer model that would distinguish it from classical BHs, as well as from some other proposed models (see below).

It is of no coincidence  that the string coupling $g_s$  determines the new time (or length) scales.  This is a natural outcome for the collapsed-polymer model because it formally introduces the fundamental string length $l_s$, which then represents a new scale from the perspective of an  external observer. Conversely,  a hypothetical  internal observer would view  the Planck length $l_P$  as the new scale. The string coupling  $\;g_s^2=(l_P/l_s)^{d-1}\;$ would then be the sole parameter that could maintain the democracy between the two points of view. In four spacetime dimensions, this is simply the small ratio $\;g_s^2= (l_P/l_s)^2 < 1\;$. The string coupling can then be viewed as the polymer's ``dimensionless $\hbar$'', $\;g_s^2 \sim \hbar G_N/l_s^2\;$.

When  it comes to theories of modified  gravity like massive gravity, the
frequencies of the new modes tend to be larger and the damping times tend to be smaller than their counterparts in GR
({\em e.g.}, \cite{massive1}), contrary to what is found here.
(Although our basic trends do happen to agree with the quasibound-state modes of these same massive-gravity models, {\em e.g.}, \cite{massive2}.)
Nevertheless,
a parametrically longer damping time was also found by the authors of \cite{Cardoso} (see also \cite{Cardoso:2016oxy,noncardoso,Abedi:2017isz,barcelo}) in a related context. Their model is based on the modes being trapped in the inner light ring of a wormhole spacetime,  and it is  meant to be representative of all  BH-like models which are not in possession of a classical BH horizon.

Their enhancement factor for the damping time (with respect to the longest-lived GR mode) scales with a certain power of a log of the ratio between the separation of the wormhole throat from $R_s$ and $R_s$ \cite{Maggio:2017ivp}. Since the exponent is much larger than unity, the scaling effectively follows a power law. On the other hand, our collapsed-polymer model introduces a new length scale $l_s$  and
includes an outer surface that acts just like a classical BH horizon when the dimensionless $\hbar$ limits to zero, $\;g_s^2\to 0\;$ \cite{emerge}. Yet, we find a power law enhancement in the damping time, similar to the findings in \cite{Cardoso,Maggio:2017ivp}.

Based on how  the  QNM  amplitudes, frequencies and damping times  scale with respect to $g_s$ for the polymer model, we are able to use  data from GW150914 to derive bounds on the string coupling.  This current observation already allows us to probe the string coupling scale in a regime which is close to that  predicted by the unification of the gravitational and gauge-theory couplings. Since the $g_s^4$ scaling in the amplitude is somewhat uncertain,  we also derive bounds on the amplitude of the polymer QNMs without assuming such a scaling. We also discuss how the bounds will improve once Advanced LIGO (aLIGO) achieves its design sensitivity.

A couple of  final notes: First, since our motivation is to  learn about actual astrophysical BHs, we will consider three large, spacelike dimensions ($d=3$)  in mind. Nonetheless, our expectation is that the basic conclusions will persist
for any  $\;d>3\;$.

Second, we are limiting considerations to Schwarzschild BHs, even though rotating Kerr BHs are more realistic. Nevertheless, as long as a Kerr BH is not  too close to extremality, the effects  of its  rotation
on the  QNM spectrum of interest  should be limited to just subdominant
corrections.

Third, a  recent complementary paper \cite{spinny} (which does consider rotating BHs)  discusses how a certain class of fluid modes, the Rossby or $r$-modes,
can be used to distinguish classical BHs from any BH-like
object that is capable of supporting fluid waves. The proposal there does not, however, discriminate between different BH-like objects.

\subsection{Organization}

The rest of the paper is organized as follows: In Sec.~\ref{sec:QNM}, the Klein--Gordon equation for  scalar perturbations is considered, from which the  QNM spectrum of collapsed polymers is derived.
We go on to explain  which modes are the most feasible in terms of GW ringdown observations and emphasize  how the amplitude, frequency and damping time of such modes
scale with respect to $g_s$.
Next, in Sec.~\ref{sec:GW}, we derive both existing and projected bounds on the polymer QNMs with current and future GW observations.
Our results are summarized in  Sec.~\ref{sec:conclusion}, followed by an appendix which  contains some background material on the collapsed-polymer model and  QNMs.

Before proceeding, we would like to briefly clarify what the collapsed-polymer model is and what it is not. The model arose out of an attempt to reconcile what is known about BHs, their associated paradoxes and  the principles of quantum gravity.
This led us to conclude that the BH interior
is described by a  state that must be strongly non-classical \cite{inny} --- so much so that it evades a description in terms of semiclassical geometry and, consequently, lacks a metric, field equations, action prinicple, {\em etc.}~\footnote{We also concluded that the interior has the same equation of state as a hot bath of long,  closed strings \cite{strungout}. Moreover, either of these properties seems to imply the other.}
And, if this picture seems far-fetched, Hawking (among others) has advocated that any description of the interior which is consistent with external observations is as good as any other \cite{info}. The polymer model has so far passed all such tests \cite{strungout,emerge}, whereas this paper is premised on looking for a new prediction that could be subjected to experimental verification.

\section{New quasinormal modes of the collapsed-polymer model}
\label{sec:QNM}

In general, an ultra-compact, relativistic object will produce two classes of QNMs when perturbed: fluid modes and
spacetime modes (see Sec.~A.2 for further discussion).  But not so for a classical BH: Because of its strictly opaque horizon and lack of interior matter, only the latter class is of any relevance.
Now, as shown in \cite{emerge}, the outer surface of a polymer BH behaves like a real BH horizon for all practical purposes. In the strict classical limit of
$\;\hbar=0\;$ ---  which for the polymer BH is equivalent to setting
$\;g_s^2=0\;$ --- the interior matter has no chance of escaping.  The polymer BH  should then, to very good approximation, agree with classical GR as far as the QNM spectrum of the spacetime modes is concerned. And so our objective is clear: To  calculate and then interpret the  spectrum for the fluid  modes when the object's interior is described by the collapsed-polymer model with a non-vanishing $g_s^2$.

This condition  of $\;g_s^2  > 0\;$ is  pivotal to
  ``stuff'' being able to leak out of the polymer BH in spite of its effective horizon.
If the strings are indeed interacting, there is no reason that smaller strings cannot break off from the long loops and then escape if they are close enough to the outer surface to avoid subsequent interactions. This process, being a perturbative quantum effect, is  of course suppressed.  One of the goals of this section is to determine the degree of this suppression, which can be calculated using Einstein's quadrupole formula and knowledge about the mode
frequencies.

 Our formal analysis begins with an appropriate form of the Klein--Gordon equation for the perturbation away from equilibrium of some physical quantity, such as the string entropy density,  string energy density and so on. A further condition is that the perturbations can couple to the spacetime fields in the  exterior region.  Here, it will be  sufficient to consider the Klein--Gordon equation for a massless scalar perturbation. Incorporating a non-vanishing angular momentum and/or spin would only complicate the practical calculations  without affecting the conclusions at a qualitative level.
We are not including any (possible)  corrections
to the Klein--Gordon equation due to the effects of string interactions, as these would necessarily scale as $g^2_s$ and thus represent  subleading corrections to the d'Alembert operator and induce only small corrections to the solutions.
Furthermore, we are effectively adopting an approximation that is akin to a Cowling approximation ({\em i.e.}, perturbations of the spacetime metric are assumed to be irrelevant to the fluid modes) \cite{cowling}. It is, however, argued in the second half of the Appendix that this
approximation is a consequence of the model in question  rather than a freely made choice.

\subsection{Wave equation and solutions}

It should be kept in mind that the ``job'' of the polymer  is to imitate  a  Schwarzschild  BH. It must then  be a spherically symmetric distribution of (stringy) matter  with an outermost (gyration) radius of $\;r=R_S\;$.

The model-dependent input is the index of refraction $\;n({\vec r})=c/v_{sound}({\vec r})\;$ or, equivalently, the speed of sound  $v_{sound}({\vec r})$ for the relevant medium. (We now set $\;c=1\;$ except when needed for clarity.) Given our assumption of spherical symmetry, the equation for the perturbation $\Phi(t,r)$ becomes
\be
\frac{1}{r^2}\frac{\partial^2 \left[r^2\Phi(t,r)\right]}{\partial r^2} \;-\;
\left[n(r)\right]^2\frac{\partial^2\Phi(t,r)}{\partial t^2}\;=\;0\;.
\label{KG}
\ee
Let us reemphasize that $\Phi$ is meant to represent the perturbation of a physical quantity (like the entropy density) and that a scalar field has been adopted to   simplify the presentation.
Equation~(\ref{KG}) is the Klein--Gordon equation for flat space such that
the coordinates $(t,r)$ are fiducial flat-space coordinates; essentially, labels for the constituent string bits.
This choice is unavoidable in the polymer model but, more generally, it is a consequence of the state of the BH interior having to  be  strongly non-classical if one insists on unitary evolution  \cite{density,inny,noclass}. The meaning of non-classicality
in this context  is that the interior defies a semiclassical  geometrical description. One can evade this predicament by  adopting  the  viewpoint that gravity is an emergent inertial force in flat space rather than a manifestation of the curvature of spacetime. This is  allowed by virtue of Einstein's equivalence principle.

Let us make one further simplifying assumption that $n(r)$ is  constant within the polymer. This may seem to be a rather severe simplification, but it  follows from the premise that matter should be   distributed uniformly throughout the interior of the polymer \cite{strungout}. This, in turn, follows from the saturation
of certain  holographic entropy bounds everywhere inside the polymer \cite{inny} which, itself, follows from an argument that the saturation of entropy bounds is a signal of non-classicality \cite{kaboom}.
 Now, with this additional assumption, the solutions to Eq.~(\ref{KG}) can be expressed  as spherical waves,
\be
\Phi(t,r) \;=\; C_o\frac{e^{-i\omega(t-nr)}}{r} +  C_i\frac{e^{-i\omega(t+nr)}}{r}\;, \hspace{.5in} r\;\leq\; R_S\;,
\label{intsol}
\ee
where $C_{o,i}$ are complex constants. Notice that the above solution contains both ingoing and outgoing waves. The latter is a consequence of  ``quantum leakage'',  allowing  modes to  escape  outside of  the (effective)  horizon.

 Applying the  usual boundary conditions for a standard QNM setup (which are itemized in Sec.~A.2), we know that
$\;C_i =- C_o\;$ because of the constraint
$\;\Phi=0\;$ at $\;r=0\;$.   We also know that $\Phi$  must be matched at the outer surface to the external solution ${\widetilde \Phi}$, which is that   of a purely outgoing wave,
\be
{\widetilde \Phi}(t,r) \;=\; C_e \frac{e^{-i\omega(t-r)}}{r}\;, \hspace{.5in}
r\;\geq\; R_S\;,
\label{extsol}
\ee
where $\;n=1\;$ has been used for  the external vacuum
 to reflect the fact that  massless fields should dominate the outward propagating wave
and
the Schwarzschild exterior has been ignored  because
it makes no sense to adopt the emergent-gravity picture on
one side of the surface and not on the other for the purpose of matching the  two solutions.   In any event, this distinction is inconsequential to the subsequent analysis because
the properties of interest (the frequencies and damping times) are determined only by the contents and
geometry of the interior region (see, {\em e.g.}, \cite{Chin}). In effect, the exterior is effectively traced out of
the calculation as far as the QNM spectrum is  concerned; see Sec.~A.2  for further explanation. Hence, in spite of the qualifier of $\;r\geq R_S\;$ in the previous equation, this solution is only strictly true at $\;r=R_S\;$.  The actual outgoing wave ${\widetilde \Phi}(t,r> R_S)$ can be described, from an external point of view, as a superposition of spacetime fields.  However, the detailed nature of this superposition is not needed for the problem at hand.

We then need to match the solutions~(\ref{intsol}) and~(\ref{extsol}) at the surface $\;r=R_S\;$. Since the amplitude of the solutions are  unknown and the time derivatives  must match if the solutions already match, this process   amounts to the sole  condition
\be
\left.\frac{\partial_r f}{f}\right|_{r=R_S}\;=\; \left.\frac{\partial_r {\widetilde f}}
{{\widetilde f}}\right|_{r=R_S}\;,
\ee
 where $\;\Phi(t,r)=e^{-i\omega t}f(r)/r\;$ and similarly for ${\widetilde f}$.

With the redefinition $\;\omega'=n\omega\;$, the above matching condition  translates into  $\;n=i\tan{(\omega'R_S)}\;$,
which is solved by \cite{Price}
\be
\omega'_{m}\;=\; \frac{m\pi}{2R_S}-\frac{i}{2R_S}\ln{\left(\frac{n+1}{n-1}\right)}\;,
\ee
where $m$ is any odd integer.

The physical frequencies are then given by
\be
\omega_{m}\;=\; \frac{m\pi}{2R_S\; n}-\frac{i}{2R_S\; n}\ln{\left(\frac{n+1}{n-1}\right)}\;,
\label{spectra}
\ee
with $\;m=1,3,5,\dots\;$ and it should be kept in mind that
$1/n$ is essentially a dimensionless $\hbar$ (this will become evident later).  Let us reemphasize that this fluid contribution to the
QNM spectrum of the  collapsed-polymer  BH  is {\em in addition}  to the usual spacetime contribution from
the BHs of classical GR.

We will encounter two important classes of  fluid QNMs; one for which $\;n\sim 1\;$ ({\em i.e.}, $\;v_{sound}\sim c$) and another for which $\;n\gg 1\;$   ($v_{sound}\ll c)$.  For the
$\;n\sim 1\;$ case, Eq.~(\ref{spectra})  becomes
\be
\omega_{m}\;\simeq\; \frac{m\pi}{2R_S }-\frac{i}{2R_S}\ln{\left(\frac{2}{n-1}\right)}\;.
\label{spectra2}
\ee
The logarithm in the imaginary part diverges, which is a sign of some problem for this case in the matching of the internal and external solutions. Indeed, going back to the solutions and substituting $\;n=1\;$, one can see that it is not possible to satisfy the   boundary conditions at $\;r=0\;$ and  $\;r=R_S\;$ simultaneously. As a result, the emission of waves for this class of modes is suppressed. Another way to see this is to take the above expression seriously; then
the  amplitude of the wave is  suppressed according to
$\lim_{n\to 1}(n-1)^{t/2R_S}$.
This suppression does appear to be a general property of relativistic fluid modes, especially
 relativistic pressure modes, as this  phenomenon has also been found in other models \cite{inversecowling,Kokk1,QNMBH,QNMBH2,Kokk2,gravi,Cardoso,Cardoso:2016oxy}.

When $\;n \gg 1\;$  --- which is expected for some of the modes, see below --- the imaginary part of the frequency now scales with $1/n^2$.
This can be shown by expanding  the logarithm in terms
of $1/n$ to obtain
\be
\omega_{m}\;=\; \frac{m\pi}{2R_S\; n}-i\left[\frac{1}{R_S\; n^2}
+ {\cal O}\left(\frac{1}{n^4}\right)\right]\;.
\label{spectra2a}
\ee

The conclusion is that the sub-relativistic modes can couple to the outer spacetime, leaking out at a rate that is determined by $\;\omega_I \propto v_{sound}^2/c^2\;$. Since the leakage has a quantum origin, we may also view $v_{sound}^2/c^2$ as the polymer's  dimensionless $\hbar$ (see below). The amplitude of the leaking modes is, however, similar in magnitude to their amplitude inside the horizon, $\;|C_e|^2\simeq |C_o|^2\;$, as a complete matching process reveals. The above conclusion applies to any partially open, spherically symmetric, very massive system with a uniform index of refraction. The   only remaining issue is  to identify the velocity of sound for the various sub-relativistic modes.

\subsection{Sound velocities in the collapsed polymer}

For the collapsed-polymer  model, one encounters a number of different mode classes according to the polymer's (or string theory's) hierarchy of parameters.
How this comes about is the next topic of discussion.

In general terms, each fluid mode can be attributed to a particular restoring force which  can act on a deformed
element of fluid. As such, the sound velocity of a mode
is determined by
\be
(v_{sound}^2)_I \;= \;\frac{K_I}{\rho}\;,
\ee
where $\rho$ is the energy density and $K_I$ is the elastic modulus corresponding to modes of type $I$. Different types include pressure modes, bending modes, shear modes, fracture modes, {\em etc}. The moduli $K_I$ have dimensions of energy density and scale as
$\;K_I\sim \frac{f_I/A} {\Delta L/L}=\frac{f_I L}{A\Delta L}\;$, where $f_I/A$ is the corresponding  force per unit  area and $\Delta L/L$ is the fractional deformation.

Let us recall that a force can  be obtained from the derivative of a free energy $F$  with respect to some geometric quantity having a  dimension of length. It follows that  each  modulus $K_I$  can be interpreted as a correction to the free energy
per unit volume $\Delta F_I/V$.
In other words,
\be
K_I \;= \;\frac{\Delta F_I}{V}\;,
\ee
and then
\be
(v_{sound}^2)_I \;=\; \frac{\Delta F_I}{V\rho}\;=\;\frac{\Delta F_I}{E}\;,
\label{sound}
\ee
where $E$ ïs the energy any $\Delta F_I$ should  be regarded as non-negative.

Let us now apply Eq.~(\ref{sound}) to  the collapsed-polymer model.
Like most any physical quantity in a string theory, the contributions to the
polymer's  free energy  can be sorted out as an expansion in both the Regge slope $\alpha'$ and the string coupling $g_s^2$ except that, in the language of the polymer model,
$\;\epsilon=l_s/R_S\;$ inherits the role of $\alpha'$.~\footnote{We subsequently work in $\;l_s=1\;$ string units.}
Importantly, the condition $\;\epsilon\ll g_s^2\ll 1\;$ is required
for the self-consistency of the polymer model \cite{strungout}.
 (For a more sophisticated explanation of how all this works, see Sec.~A.1 and Eqs.~(\ref{expansion1}),~(\ref{expansion}) in particular.)

 Identifying each order in the
expansion as the correction due to a different mode, we can write
the leading correction   to the  ``tree-level''
free energy  $\;F_0 \sim  M_{BH}\;$ ($M_{BH}$ is the
polymer/BH mass)
  as
$\;\Delta F_1 \sim g_s^2 F_0\;$. Then $\;\Delta F_2 \sim \epsilon F_0\;$,  $\;\Delta F_3 \sim g_s^4 F_0\;$ and so on.  Each of the corrections, including the zeroth-order term,
can be expected to correspond to some independent class of modes; some examples  are discussed below.

The speed of sound
in the stringy interior
can be read off of Eq.~(\ref{sound}) for any of the modes.
For instance, since  $\;F_0\approx M_{BH}=E\;$, the corresponding mode
is a relativistic wave, $\;v_{sound}=1\;$.  The pressure ($p$) modes, which are associated with volume deformations of the interior, are an example of relativistic waves. This conclusion is based on the observation  that $\;p=\rho\;$ for a highly excited state of closed strings; this is a well-known result \cite{AW} and  also follows from Eq.~(\ref{pressure}) in the Appendix. Consequently, the bulk modulus for the polymer  is $\;K_B=\rho\frac{dp}{d\rho}=\rho\;$, from which $\;(v^2_{sound})_B= K_B/\rho=1\;$  follows.
To sum up, the pressure modes and  their analogues  are based on  leading-order changes to the effective free energy and have a speed of sound of $\;v_{sound}= 1\;$. As  argued
earlier, such relativistic modes effectively decouple from the outer region of spacetime and cannot be used to probe the inner structure of the BH.

A  more interesting class of modes  is that for which the free-energy  correction scales as $\;\Delta F_1 \sim g_s^2 F_0\;$; these being the leading-order non-relativistic modes.  For this class, $\;v_{sound}^2 = g_s^2 c^2\;$ and the  frequency of emitted GWs then scales as $\;\omega \sim v_{sound}/R_S \sim g_s\, c/R_S\;$, whereas  the damping time due to mode leakage to the outside scales as $\;\tau_{damp} \sim (1/g_s^2)\, (R_S/c)\;$ as follows from Eq.~(\ref{spectra2a}). Here, one can see  explicitly that $\;g_s^2=v_{sound}^2/c^2\;$; and so  both of our estimates for  the dimensionless $\hbar$ coincide, with one coming from  first principles
(see Sec.~1.2) and another by estimating  the amount of leakage from the horizon (see Sec.~2.1).

By counting powers of the coupling $g^2_s$ and powers of the number of string ``bits'' $N$ ($N=S_{BH}\sim M_{BH}/\epsilon$) in the free-energy correction $\;\Delta F_1\sim g_s^2N\epsilon \;$, one can  attribute this class
of modes to   the splitting and subsequent rejoining interactions of single loops of  strings. The reasoning behind this claim is that
 each splitting has a free energy ``cost" of $g_s^2$, as does each subsequent rejoining.
Meanwhile, the single factor of
$N$  implies  that only a single string loop can  be involved in any one  interaction (as the typical length of a string loop is of order $N$ in string units \cite{SS,LT}).
A physical example from this class  is a fracture mode, whereby  a ``crack-like" defect propagates in the stringy material due to the continual
splitting and rejoining of strings.

Other, higher-order classes of modes are less interesting because they are associated with  extremely non-relativistic speeds of sounds (recall that $\;\epsilon\ll g^2_s$), rendering the frequencies too slow to be relevant in any realistic  situation. Nevertheless, it is still interesting to ask about the physical meaning of these classes. For example, those associated with $\;\Delta F  \sim \epsilon^2 F_0\;$ would include bending modes. This is because the (free) energy ``cost" for  bending scales as the spacetime curvature, $ \;\Delta F_{bend}\sim F_0/R_S^2 \sim F_0 \epsilon^2\;$. In a sense, these modes also decouple from the exterior
but for a different reason than the pressure modes.

All classes of modes are also subject to intrinsic dissipation. To estimate the strength of this dissipation, we will assume that it  is caused by the shear viscosity $\eta$. This is because we have a good understanding of the scaling properties of the shear viscosity for the collapsed-polymer model in particular and for  BH-like objects in general. Let us start here with an appropriate expression for
the rate of intrinsic dissipation $\;1/{\widetilde \tau}\;$ \cite{instability},
\be
\frac{1}{\widetilde\tau}\;=\; (\ell-1)(2 \ell+1)\int_0^{R_S} dr \, r^{2 \ell} \eta \left( \int_0^{R_S} dr \rho r^{2\ell+2} \right)^{-1}\;,
\label{diss}
\ee
where $\ell$ is the angular momentum of the mode.

In the case of the  polymer model --- for  which the stringy matter  saturates the so-called KSS bound \cite{KSS} throughout
the interior \cite{emerge} --- the relevant expressions are
$\;\rho=1/(g^2_s r^2)\;$ and
  $\;\eta=s/(4\pi)= 1/(4\pi g_s^2 r)\;$ \cite{strungout}, where $s$ is the entropy density.  Substituting these
into Eq.~(\ref{diss}),  we then have
\be
\frac{1}{\widetilde\tau}\;=\; (\ell-1)(2\ell+1)  \int_0^{R_S} dr \frac{1}{4\pi} r^{2\ell -1} \left( \int_0^{R_S} dr r^{2\ell} \right)^{-1}\;=\;\frac{1}{4\pi} \frac{(\ell-1)(2 \ell+1)^2}{2 \ell} \frac{c}{R_S}\;.
\ee
Restricting to  the choice $\;\ell=2\;$, as is most relevant to GW production,
we finally  obtain
\be
\frac{1}{\widetilde\tau}\;=\; \frac{25}{16\pi} \frac{c}{R_S}\;.
\label{tauV}
\ee

The result in Eq.~(\ref{tauV}) applies to relativistic modes. For non-relativistic modes, the ratio  $\eta/\rho$  scales with  $\;(v_{sound}/c)^2$. This behavior can be
understood by starting with the diffusion equation for viscous flow ---  for which $\eta/\rho$
serves as  the diffusion coefficient ---  and then
making the sound velocity equal to $c$ with the  rescaling $\;r\to  (v_{sound}/c)r\;$.
 It then  follows that
\be
\frac{1}{\widetilde\tau}\;=\; \frac{25}{16\pi}\frac{v^2_{sound}}{c^2} \frac{c}{R_S}\;
\ee
or, for the fracture modes in particular,
\be
{\widetilde\tau}\;=\;2 \frac{1}{g_s^2} \frac{R_S}{c}\;\simeq\;2\tau_{damp}\;.
\ee
And so the time scale for intrinsic dissipation is comparable to that of  damping.

To summarize,  the relativistic  modes are ``unaware'' of the existence of
any new physical scale, whereas  the fracture modes and their analogues would
present a tell-tale distinction. This contrast can be attributed to the introduction of the string-coupling scale ---
the ratio $l_P/l_s$ ---   as its inclusion  modifies the spectrum of the fracture modes in a substantial way.
The implication being  that the QNM spectrum  of  a collapsed polymer has a  definitive and potentially  observable signature.

\subsection{Estimate of gravitational-wave emission from polymer black holes}
\label{sec:QNM-amp}

The goal of this subsection is  to estimate  the  relative amplitudes of the emitted GWs
and then  compare
the fracture-mode amplitudes with those due to the spacetime modes. The quadrupole formula can be used to obtain the desired ratio of amplitudes since we
know about  the respective  energies and frequencies of the emitted waves.
It should be emphasized that the amplitudes, as estimated here,  are much less certain than the frequencies and damping times.

Let us first recall that the (free) energy of a fracture mode scales as $\;E_{frac}\sim g_s^2 F_0 \sim g_s^2 M_{BH}\;$, whereas the energy in a GW corresponding to  a  spacetime  mode scales as $\;E_{st}\sim M_{BH}\;$.
The ratio of energies then scales as
\be
\frac{E_{frac}}{E_{st}}\;\sim \; g_s^2 \;,
\label{Eratio}
\ee
where $E_{st}$ can be estimated via observations; for example, in GW150914, GWs carried away  about 5\% of the total mass of the merging BHs.
Let us also recall that the ratio of their squared frequencies
scales in the same way,
$\;\omega_{frac}^2/\omega_{st}^2\simeq g_s^2\;$.

Now, using  the quadrupole formula to estimate   the GW strain amplitude $h$,
one finds  that the relative amplitudes of the emitted GWs scale  according to
\be
\frac{h_{frac}}{h_{st}} \;\simeq \;\frac{E_{frac}}{E_{st}} \frac{\omega_{frac}^2}{\omega_{st}^2} \;\simeq \;(g_s^2)^2
\;,
\label{amps}
\ee
where $\;Q \sim E\, R_S^2\;$ has been used  to estimate the \emph{fraction} of the quadrupole moment that contributes to the GW production for each mode.
The parameter $g_s$ is expected to be small, but not extremely small, as explained previously.

If the string coupling is indeed not too small, one can anticipate some spectacular observational consequences. For concreteness, let us set $\;g_s^2=1/10\;$ and choose  the other  parameters to be those of GW150914 --- meaning an
observed ringdown of   $\;f=251~{\rm Hz}\;$ and a damping time (in addition
to the standard ringdown time  $\;1/2\pi f\simeq 0.6~{\rm ms}$)
of  $\;\tau = 4~ {\rm ms}\;$  \cite{LIGO1}.
The new class of GWs  are  reduced  in amplitude by a factor  of about $1/100$ in comparison to those already observed but oscillate with  frequencies about three times lower, $\;\omega \sim 2\pi(251~{\rm Hz})/3 \sim 500~{\rm Hz}\;$,
and have damping times which last  about  ten times longer,
$\;\tau \sim 40~{\rm ms}\;$.
Because of their lengthier ringdown time, the sensitivity for detection of the new class of GWs, as estimated by $h/\sqrt{\mathrm{Hz}}$, is enhanced by a factor of  $\;\sqrt{\tau_{damp}}\sim g_s\;$.~\footnote{This enhancement follows
from two competing effects: The opportunity for signal detection increasing linearly with  time versus the noise increasing only as $\sqrt{t}$. Here, the relevant time scale is the ringdown time.}  This means that the sensitivity for detection  has decreased by ``only'' a factor of $g_s^3$, rather than the factor $g_s^4$ as estimated above.
Such a $g_s^3$ scaling in the signal-to-noise ratio (SNR) will be confirmed
in the following section.

\section{Bounds on polymer modes from gravitational-wave observations}
\label{sec:GW}

We will start off in this section by  using the events  GW150914 and GW151226 to  derive current bounds on the  polymer modes. Following this,  future projected bounds that are based on  the aLIGO design sensitivity will also  be derived.  A subscript of $p$ or $BH$ is used to distinguish between properties of the polymer modes and classical BH modes respectively.

\subsection{Gravitational-wave spectrum and signal-to-noise ratio}

Let us begin here by representing the polymer QNMs as damped, sinusoidal waveforms,
\be
h(t)\;=\; A_p\, e^{-(t-t_p)/\tau_p} \sin [2 \pi f_p (t-t_p) - \phi_p] \, \Theta (t-t_p)\;,
\ee
where $\Theta$ is the Heaviside step function, $A$ is a QNM amplitude,
$f$ is a QNM frequency and $\tau$ is a QNM damping time. Also, $t_p$ is the time delay of the  polymer QNM  relative  to that of a typical  GR mode and $\phi_p$ is a constant phase. The time delay $\;t_p\sim 1/\omega_p\sim 1/g_s \omega_{BH}\sim \tau_{BH}/g_s\;$ ensures that these and the classical GR modes will not be superimposed
to any significant degree, although
this detectability  may  deteriorate if one includes the fundamental mode due to possible degeneracies among parameters.
The Fourier transform of the above equation works out to be
\bea
\tilde h(f) &=& e^{2\pi i f t_p} A_p \tau_p \frac{2f_p^2 Q_p \cos \phi_p- f_p (f_p-2 i f Q_p) \sin \phi_p}{f_p^2-4i f f_p Q_p + 4 (f_p^2-f^2)Q_p^2}\;,
\eea
with  $\;Q_p \equiv \pi f_p \tau_p\;$. The above expression reduces to Eq.~(2.2) of \cite{Berti:2007zu} when $\;t_p = 0\;$. Notice as well  that $|\tilde h|$ does not depend on $t_p$.

To assist in estimating $A_p$, we will use
$\;A_p \sim g_s^4 A_\BH\;$ ({\em cf}, Eq.~(\ref{amps})) and thus require the  amplitude of the   QNMs from a  classical  BH \cite{Berti:2005ys},
\be
\label{eq:BH-ringdown-amp}
A_\BH \;=\; \frac{M_\BH}{r} \mathcal{F} \sqrt{\frac{8 \epsilon_\mrm{rd}}{M_\BH Q_\BH f_\BH}}\;,
\ee
where $r$ is the distance to the source, $\mathcal{F}$ is a function that depends on the source location, $\epsilon_\mrm{rd}$  is the ringdown  efficiency and $\;Q_\BH \equiv \pi f_\BH \tau_\BH\;$. The fitting formula for $f_\BH$ and $Q_\BH$ of a BH forming in the aftermath of  a binary coalescence of BHs is given in \cite{Berti:2005ys}. Here, we are setting the spins of the initial BHs to zero for simplicity.
The efficiency is roughly given by $\;\epsilon_\mrm{rd} \approx 0.44 q^2\;$ for non-spinning BH binaries \cite{Berti:2007fi}, where $\;q \equiv m_1 m_2/(m_1+m_2)^2\;$ is the symmetric mass ratio of a binary with individual masses $m_1$ and $m_2$.

Let us  now estimate the SNR of collapsed polymers by  using \cite{cutlerflanagan}
\be
\label{eq:SNR}
\mathrm{SNR}^2  \;=\; 4 \int_{f_\mrm{min}}^{f_\mrm{max}} \frac{|\tilde h (f)|^2}{S_n (f)} df\;,
\ee
where  $f_\mrm{min}$ and $f_\mrm{max}$ are the minimum and maximum frequency ---
for which we choose the values  $\;f_\mrm{min}=10~{\rm Hz}\;$,
$\;f_\mrm{max} = 3000~{\rm Hz}\;$ unless otherwise stated ---  whereas
 $S_n$ is the detector's  noise spectral density.  The density $S_n$ for the
aLIGO O1 run
 is given by \cite{TheLIGOScientific:2016zmo} and the fit can be found  in Appendix C of \cite{pretorius}, while  that for  aLIGO's design sensitivity with the zero-detuned, high-power configuration is given in  \cite{Ajith:2011ec}.

\begin{figure}[htb]
\centerline{\includegraphics[width=11.5cm]{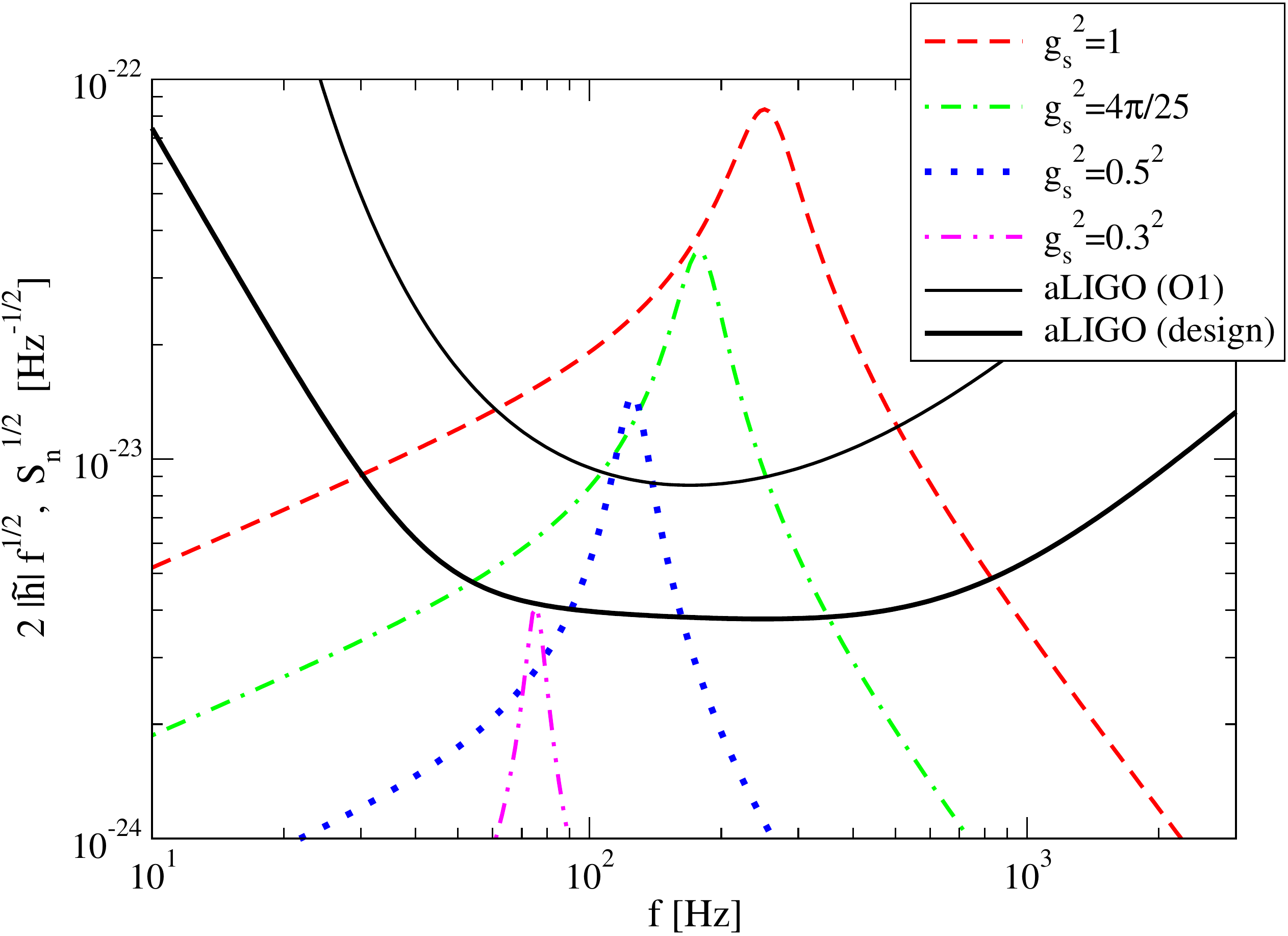}}
\caption{QNM spectrum of putative polymer modes for GW150914 with various $g_s^2$, as well as  noise spectral density against frequency. For $g_s^2=1$,  the QNM amplitude, frequency and damping time for the polymer modes  are the same as those of a classical BH. The ratio between the signal and noise roughly corresponds to the SNR. The spectrum is detectable if this  ratio is above the threshold $(\sim 5)$.}
\label{fig:spectrum}
\end{figure}

Figure~\ref{fig:spectrum} compares the noise spectral density to the polymer QNM spectrum for various values of  $g_s^2$, with the other  parameters chosen to be consistent with GW150914 ($m_1 = 35.7~M_\odot\;$, $\;m_2 = 29.1~M_\odot\;$,
$\;r=410~{\rm Mpc}\;$, $\;f_\BH = 251~{\rm Hz}\;$,
$\;\tau_\BH = 4~{\rm ms}\;$~\cite{LIGO,LIGO1}). We have used the scaling relations
$\;A_p \sim g_s^4 A_\BH\;$, $\;f_p \sim g_s f_\BH\;$, $\;\tau_p \sim \tau_\BH/g_s^2\;$ as
motivated  in the previous section and set  $\;\phi_p = 0\;$ for simplicity.
The value of  $\mathcal{F}$  in Eq.~\eqref{eq:BH-ringdown-amp} is chosen
by requiring   that the SNR equals 7 for the case of a  classical BH   with GW150914 parameters \cite{pretorius,Berti:2016lat} (and with
$\;f_\mrm{min}=222~{\rm Hz}\;$ in Eq.~\eqref{eq:SNR}, which  corresponds to the frequency where the spectrum peaks \cite{pretorius}). We have plotted  $2 |\tilde h | \sqrt{f}$ instead of $|\tilde h|$ for the signal spectrum so that the ratio between the signal and noise in Fig.~\ref{fig:spectrum} goes roughly as the SNR
({\em cf}, Eq.~\eqref{eq:SNR}). Notice that the spectrum's  amplitude and width both grow larger as one increases $g_s^2$.

\subsection{Current and future bounds with gravitational-wave observations}

Continuing with the same setup as in the previous subsection, we will next use GW observations to  derive bounds on the polymer modes. It will initially be assumed
that  the QNM amplitude scales with  $g_s^4$
as  explained in Sec.~\ref{sec:QNM-amp}; however,  this assumption
will be  relaxed later on.

\begin{figure}[htb]
\centerline{\includegraphics[width=9.cm]{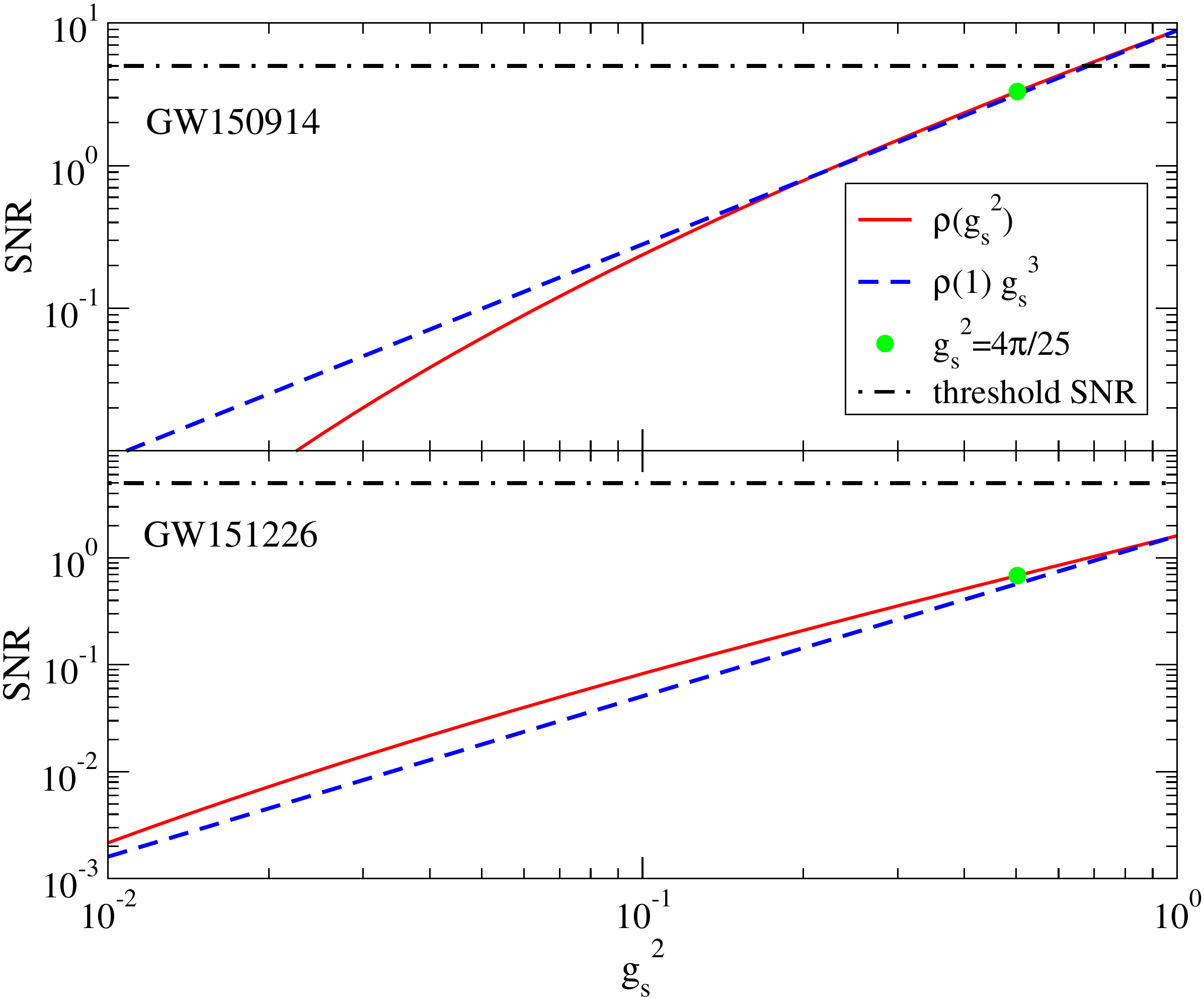}}
\caption{{\bf Top:} SNR of the putative QNM of a collapsed polymer for GW150914 as a function of $g_s^2$ (red, solid). The SNR scales with $g_s^3$ (blue, dashed) for $g_s^2 \sim 1$ as explained in Sec.~\ref{sec:QNM-amp}. The SNR threshold of 5 (black, dotted--dashed) allows us to constrain $g_s^2$ as $g_s^2 \lesssim 0.65$. As the detector sensitivity increases, one will be able to probe $g_s^2$ for the unification of the gravitational and gauge theory couplings (green dot). {\bf Bottom:} Same as the top panel but for GW151226.}
\label{fig:SNR}
\end{figure}

\subsubsection{Bounds assuming the $g_s^4$ amplitude scaling}

The top panel of Fig.~\ref{fig:SNR} presents the SNR for the QNMs of a collapsed polymer with  GW150914 parameters. We have  used two aLIGO detectors (corresponding to Hanford and Livingston) with the O1 run. For $\;g_s^2 \sim 1\;$,
the SNR scales with $g_s^3$ as discussed at the end of Sec.~\ref{sec:QNM-amp}. This scaling is valid for a white-noise background; however,  as  $g_s^2$ becomes smaller, there is an extra suppression due to the frequency dependence of the noise curve. Namely, as one lowers $g_s^2$, the QNM frequency $f_p$ becomes smaller and enters a range where  the detectors are less sensitive (see Fig.~\ref{fig:spectrum}).

The bottom panel of Fig.~\ref{fig:SNR} depicts the SNR for the case of a collapsed polymer with GW151226 parameters. The value of  $\mathcal{F}$ in Eq.~\eqref{eq:BH-ringdown-amp} is now chosen by requiring that the SNR equals unity for a classical BH with  GW151226 parameters. The predicted $f_\BH$ for this source is $\sim 790$~Hz, which is higher than the corresponding frequency in the previous case (251~Hz for GW150914). Consequently,  as $g_s^2$ becomes smaller, $f_p$ is actually entering the region where the detector is  {\em most} sensitive. Meaning that  the scaling of the SNR with $g_s$ is shallower than $g_s^3$.

Let us now derive an upper bound on $g_s^2$ by using the knowledge  that the LIGO--Virgo Collaboration did not report the presence of an additional ringdown signal on top of the dominant BH signal.~\footnote{References \cite{noncardoso,Abedi:2017isz} reported the presence of ``echoes'' on top of the primary ringdown signal. This claim  is apparently  still in debate \cite{Ashton:2016xff} as the result has not yet been confirmed by other groups.} This means that we can  derive bounds on the polymer modes under the assumption that the observed data is consistent with gravitational waveforms from binary BH mergers in classical GR. It then follows that the SNR for the polymer modes has to be smaller than the threshold value. For example, if the threshold is 5~\cite{Bhagwat:2016ntk,Yang:2017zxs} --- as indicated  by the horizontal, black, dotted--dashed line in the top panel of  Fig.~\ref{fig:SNR} ---  one can use GW150914 to roughly bound  $g_s^2$ such that   $\;g_s^2 \lesssim 0.65\;$ (the upper limit being where the red, solid curve crosses the black, dotted--dashed line). This upper bound is intriguingly close to the point where $g_s^2$ corresponds to the unification of the gravitational and gauge coupling constants,  $\;g_s^2 = 4\pi/25 \sim 0.5\;$.

\begin{figure}[htb]
\centerline{\includegraphics[width=9.cm]{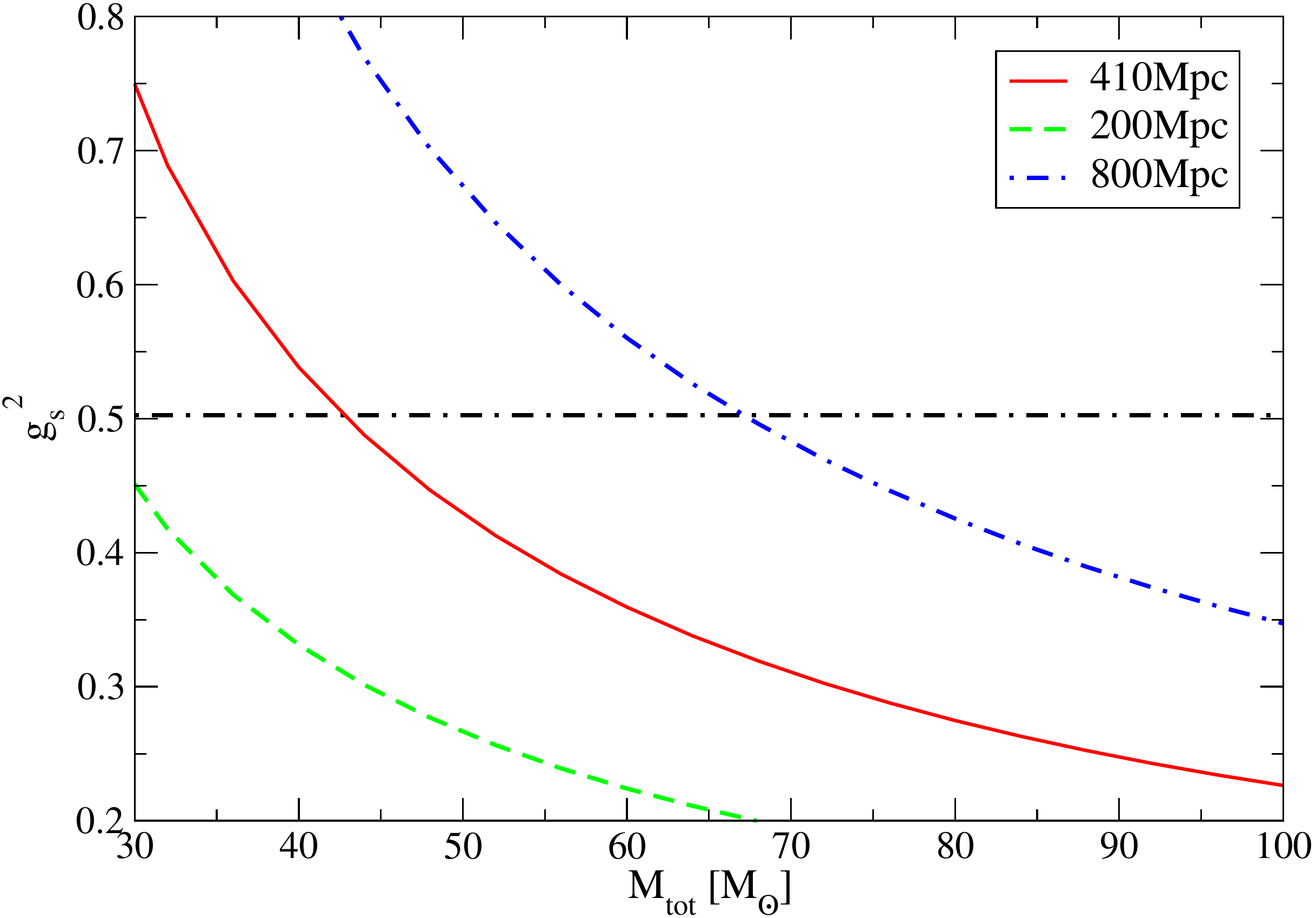}}
\caption{Projected upper bound on $g_s^2$ as a function of the total mass of an equal-mass BH binary at various distances apart using aLIGO's design sensitivity. 410~Mpc corresponds to the distance for GW150914 \cite{LIGO}. The horizontal line represents $g_s^2 = 4 \pi/25$. The upper bound on $g_s^2$ scales with $r^{2/3}$.}
\label{fig:gs2-bound}
\end{figure}

It is also interesting to consider the  future prospects for  constraining $g_s^2$ with GW observations. Figure~\ref{fig:gs2-bound} displays  the projected upper bound on $g_s^2$ given  aLIGO's design sensitivity (again using the two interferometers at Hanford and Livingston) and  assuming that aLIGO does not find the collapsed polymer signal. In other words, such an upper bound is equivalent to the \emph{minimum} $g_s^2$  for which  aLIGO would be able to detect such a signal.
We have, for  concreteness,  used the sky-averaged value of $\mathcal{F}$ in Eq.~\eqref{eq:BH-ringdown-amp}, assumed that the initial binary contains equal-mass BHs at various distances $r$ apart and adopted a threshold SNR of 5. As evident from the figure, one can constrain $\;g_s^2 \lesssim 4\pi/25\;$ for a total mass of $45$~$M_\odot$ or larger when $\;r=410\;$~Mpc. Given that   $\;\mathrm{SNR} \propto g_s^3/r\;$ and that  $g_s^2$ is  determined by the SNR being equal to its threshold value, one finds that such an upper bound on $g_s^2$ is proportional to $r^{2/3}$. We have checked that this analytic scaling in distance agrees with the displayed results in Fig.~\ref{fig:gs2-bound}.

The bounds on $g_s^2$  will further increase as (i) the number of interferometers increases, (ii) the detector sensitivity improves and (iii) one is able to combine signals from multiple sources.
We stress that the upper bounds presented here are not robust and should be understood as only rough estimates.

\subsubsection{Bounds without assuming the $g_s^4$ amplitude scaling}

Since the $g_s$ scaling in $A_p$ is the most uncertain among $A_p$, $f_p$ and $\tau_p$, it is perhaps more  appropriate to   place a bound on the relative amplitude $\;\gamma \equiv A_p/A_\BH\;$ without assuming the scaling $\;A_p \sim A_\BH g_s^4\;$. Let us first work out a simple scaling relation for the upper bound on $\gamma$. We start with  $\;\mathrm{SNR} \propto A_p \sqrt{\tau_p} \propto \gamma A_\BH/g_s\;$  and then, like before, require this  SNR to be equal to its threshold value.
On this basis, one finds that the upper bound on $\gamma$ scales linearly with $g_s$.

\begin{figure}[htb]
\centerline{\includegraphics[width=9.cm]{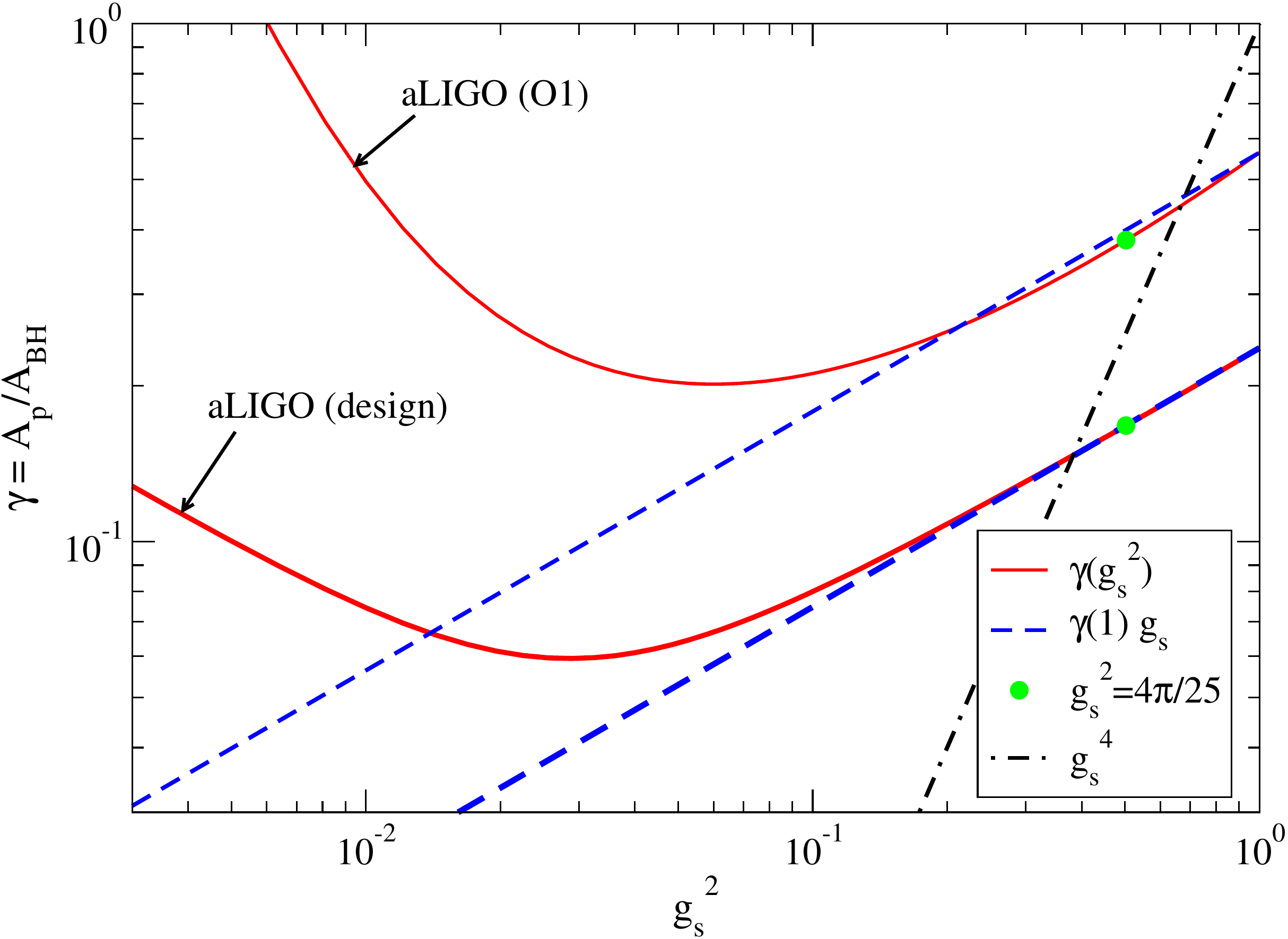}}
\caption{Upper bound on the relative amplitude $\gamma$ of the polymer QNMs with respect to the BH QNMs as a function of $g_s^2$. The thinner red, solid curve is the current bound from GW150914 with aLIGO's O1 run, while the thicker red, solid curve is the projected bound for a GW150914-like event with aLIGO's design sensitivity. Blue, dashed lines are the analytic prediction for the upper bound on $\gamma$ valid around $g_s^2 =1$, while green dots are the bounds at $g_s^2 = 4\pi/25$. The black, dotted-dashed line is the predicted relative amplitude proportional to the $g_s^4$ scaling (see Sec.~\ref{sec:QNM-amp}).}
\label{fig:rel-amp}
\end{figure}

The thinner red, solid curve in Fig.~\ref{fig:rel-amp} shows the upper bound on $\gamma$ from GW150914 with the threshold SNR of 5. These results roughly agree with the linear scaling in $g_s$, as motivated above, when $\;g_s^2 \sim 1\;$
({\em cf}, the uppermost dashed, blue line). The slight deviation from the linear scaling in this regime can be attributed  to a small frequency dependence in $S_n$ around $\;f=f_p\;$.  On the other hand, the curve strongly  deviates away from the linear scaling when $\;g_s^2 \ll 1\;$. This is because the spectrum falls  out of the detector's frequency band as one decreases $g_s^2$. The GW150914 observation sets $\;\gamma \lesssim 0.38\;$ for $\;g_s^2 = 4\pi/25\;$. But, if $\;\gamma = g_s^4\;$ (see the black, dotted--dashed line) as predicted in Sec.~\ref{sec:QNM-amp},
then the SNR of the polymer modes for GW150914 becomes smaller than the threshold already when $g_s^2 \lesssim 0.65$ ---  in agreement  with the top panel of Fig.~\ref{fig:SNR}.

The thicker red, solid curve in Fig.~\ref{fig:rel-amp} depicts the projected bound on $\gamma$ when using the noise curve for aLIGO's design sensitivity. One should
first observe  that the linear-in-$g_s$ scaling near $\;g_s^2 \sim 1\;$ is a better fit than that found for aLIGO's O1 run because the noise curve is flatter for the future design sensitivity (see Fig.~\ref{fig:spectrum}). Second,  the upper bound on $\gamma$ decreases by a factor of $\sim 2$ at $\;g_s^2 = 4\pi/25\;$ in
comparison to    the current bound from GW150914.

\section{Conclusion}
\label{sec:conclusion}

We have discussed how the interior structure of BHs,  as described by the collapsed-polymer model, affects the spectrum of QNMs. Our main result is the identification of several new  classes of  QNMs, in addition to the classical GR modes
which are a common   feature of all  BH-like objects with an effective horizon
or a light ring~\cite{Cardoso,Cardoso:2016oxy,Cardoso:2014sna,Barausse:2014pra,Barausse:2014tra}.  We  found  sub-relativistic modes whose sound velocity is $\;v_{sound} \simeq g_s c\;$;  these being associated with the  self-interactions of the strings.  Additionally, there are many other  classes of  exceptionally slow modes
that are induced  by weak restoring forces; for instance, one such class describes bending modes with a sound velocity
of $\;v_{sound}\simeq c\;l_s/R_S \;$.

We have also discussed how the new classes of QNMs could affect the emission of GWs from BHs.  The emission due to relativistic modes is  suppressed to such an extent that they essentially decouple from the outer spacetime --- in agreement with previous studies in the literature on fluid modes in ultra-compact objects. The various classes of exceptionally slow modes are irrelevant because  their low frequencies necessitate prohibitively long observation times.  Fortunately, the emission due to the  leading-order sub-relativistic modes was shown  to lead to an interesting observable signature: A characteristic ringdown by the emission of low-frequency GWs which follow the conventional emissions after a relatively brief but distinguishable time delay. The amplitude of this new class of GWs is lower than the amplitude of the usual BH GWs  by a factor $(g_s^2)^2$.

Our main conclusion is   that  observations of GWs from colliding BHs provides a means for   differentiating the collapsed-polymer model  from the BHs of classical GR.
These distinctions --- the lower frequencies and time delay --- are determined mainly by the string  coupling, which itself depends on the ratio of the Planck scale to the  string  scale and  is also the dimensionless $\hbar$ for the
polymer. Remarkably, we found that GW150914 places an upper bound on $g_s^2$ that is close to $4\pi/25$, and such a bound will only become stronger as the detector sensitivity improves.

One may still wonder how the fluid modes appear to evade the BH horizon   as seen from an external, asymptotic observer's perspective. After all,
a horizon must be there as far as {\em this observer} is concerned, regardless of whether it is a classical BH or merely a
BH-like object with an effective horizon. This is an important question in its own right and will be addressed in a separate discussion \cite{ridethewave}.

\section*{Acknowledgments}

We thank Frans Pretorius and Ofek Birnholtz for discussions on the observability of the emitted GW. We also thank John Boguta, Roman Konoplya and Paolo Pani for useful comments on the manuscript.
The research of RB was supported by the Israel Science Foundation grant no. 1294/16. The research of AJMM received support from an NRF Incentive Funding Grant 85353 and an NRF Competitive Programme Grant 93595. AJMM thanks Ben Gurion University for their  hospitality during his visit.
K.Y.~acknowledges support from JSPS Postdoctoral Fellowships for Research Abroad, NSF grant PHY-1305682 and Simons Foundation.

\appendix

\section{A brief review on background\label{sec:back}}

\subsection{The collapsed polymer}

The polymer model  assumes
that the BH interior consists of a hot bath of closed, interacting strings in a finite volume. The properties of
such  a system are explained in  \cite{SS,LT} (also see \cite{strungout,emerge}.)

Let us start here  by considering a free, highly excited, closed string of  length $L$ in an infinite space. In this case, the string occupies a region whose  linear size $R$ is given by the random-walk scale, $\;R\sim l_s\sqrt{L/l_s}\;$. One can  regard $\;N=L/l_s\;$ as the total  number of ``string bits" in the state, and so  $\;R\sim l_s \sqrt{N}\;$. The situation, however, changes when strings interact, which they do  by splitting and joining. Such interactions induce an effective attraction that causes the strings to occupy a smaller region in space, leading to a smaller value of $R$  \cite{HP,DV}. Since the only relevant scales
are $l_s$ and $l_P$ and the strings do not ``know'' about
the latter,  one expects that $\;R\sim l_s N^\nu\;$ for some $\nu$ which could be different than $1/2$. The resulting picture is a finite-sized, bound state
of strings that is dominated by about  $\;\ln{N}\;$ long  loops
\cite{SS,LT}.

The parameter $N$ also measures   the entropy of the string state and, since $\;N\sim (R/l_s)^{1/\nu}\;$, the entropy will not, in general,  be extensive. An area law, as in the case of BHs,  implies that $\;\nu=1/(d-1)\;$ with $d$ being the number of spatial dimensions. A  scaling relation with entropy in terms of $R$ is also described by the Flory--Huggins theory of polymers \cite{flory}.~\footnote{See the books by De Gennes \cite{bookdg}, and Doi and Edwards \cite{bookde} as well.} This theory is reexamined in \cite{degennes} and reviewed  in, for instance, \cite{polymer}. The parameter $\nu$ is called the Flory exponent and the temperature at which the polymer becomes tensionless is known as the Flory temperature. The linear size $R$ is referred to  as the gyration radius of the polymer and $N$ represents the total number of monomers within the polymer chain(s). For our case of attractive interactions,  the gyration radius  is smaller than $l_s \sqrt{N}$ and the system is then identifiable as a ``collapsed polymer".

The  theory of collapsed polymers has been adopted to show that the bound state of highly excited strings can be described by a quadratic (effective) free energy
\cite{strungout}. In string  ($l_s=1$) units, this free energy $F$ takes the form
\be
-\left(\frac{F}{T_{Hag}}\right)_{strings}\;=\; \epsilon N - \frac{1}{2} \frac{g_s^2}{V} N^2\;,
\label{FES1}
\ee
where $g_s$ is the string coupling, $\;V\sim R^{d}\;$ is the occupied volume,
 $T_{Hag}$ is the Hagedorn temperature and we  disregard  an order-one numerical factor so that $\;T_{Hag}=1\;$ in string units. The parameter $\epsilon$ is an effective, dimensionless temperature which measures
the deviation of the actual temperature $T$ from the Hagedorn value, $\;\epsilon=(T-T_{Hag})/T_{Hag}\;$.
The equilibrium solution of the theory,
which is obtained by minimizing the free energy with respect
to $N$, enforces the relation
\be
\label{consol}
\frac{N}{V} \;=\; \frac{\epsilon}{g^2_s}\;.
\ee

The collapsed-polymer scaling relations agree with those of a BH when the
parameters of the polymer theory --- $N$, $\epsilon$ and
$g_s^2$ ---  are related to those of the BH --  the Schwarzschild radius $R_S$,  energy $M_{BH}$ and entropy $S_{BH}$ ---  in a specific way \cite{strungout}.
In particular,~\footnote{Here and for the remainder, the string length
$l_s$,  fundamental constants and order-unity numerical factors will only be made explicit when needed for clarity.}
\be
R_S \;=\;\frac{l_s}{\epsilon}\;,
\label{bhvdef}
\ee
meaning that the Hawking temperature is
\be
\label{bhthaw}
T_{Haw}\;=\;\epsilon\;.
\ee

Additionally,  the BH  entropy is
\be
\label{bhs}
S_{BH}\;=\;N\;=\; V\frac{\epsilon}{g_s^2}\;=\;
\left(\frac{R_S}{l_p}\right)^{d-1}\;,
\ee
where $l_P$ is the Planck length, the second equality follows from Eq.~(\ref{consol}) and the last
one  from $\;g_s^2=(l_P/l_s)^{d-1}\;$  as well as $\;R=R_S=1/\epsilon\;$.
Also, the total energy of the bound state is
found to be in agreement with that of the  BH
({\em cf}, Eq.~(\ref{pressure}) for the density $\rho$),
\be
\label{bhm}
E_{bound}\;=\;V\frac{\epsilon^2}{g_s^2}
\;=\;  \frac{1}{l_P}\left(\frac{R_S}{l_P}\right)^{d-2}\;=\; \epsilon N \;=\; M_{BH}\;.
\ee

It is worth noting that the pressure $p$ is equal to the
energy density $\rho$ for a highly excited state of closed strings
\cite{AW}. This equality also follows directly from the free energy~(\ref{FES1}), both at and away from equilibrium. Using standard thermodynamics,   one finds that
the equilibrium values are
\be
p\;=\; \rho \;=\; \frac{\epsilon^{2}}{g_s^2}\;.
\label{pressure}
\ee
This pressure is not to be confused with
the (effective) tension,
$\;\sigma=\frac{\partial F}{\partial L}\;$, which
 vanishes at equilibrium by virtue of $\;L=l_s N\;$.

For self-consistency, the string-theory parameters must obey
the following relations \cite{strungout}:
$\; \epsilon\ll g^2_s \ll 1\;$ and $\;g^2_s N = V\epsilon \gg 1\;$. Together, these ensure that the BH is large in string units, the coupling is small but finite and    the higher-order interaction terms in the free energy in Eq.~(\ref{FES1})
are suppressed. The higher-order terms can come from
$\;\alpha^{\prime}\sim l_s^2\;$  corrections, additional loop corrections  or their combination. Because the former is controlled by    the Regge slope   $\;\alpha^{\prime}\propto \epsilon^2\;$,
we know that the equilibrium form of the corrected
free energy  looks schematically like
\be
F\;\sim\; \epsilon N\left[1+a_1 g_s^2+ a_2 g_s^4+\cdots\right]\times\left[1+ b_1 \epsilon+ b_2 \epsilon^2+\cdots\right]\;,
\label{expansion1}
\ee
where
the odd powers of $\epsilon$ are shorthand
for  powers of  $g_s^2N/V$ and  originate from loop corrections
(the even powers of $\epsilon$ could be of either type), whereas
 the explicit  powers of $g^2_s$ are from string self-interactions.
The above hierarchy  tells us that the next-to-leading
term in the expansion has a suppression factor of $g_s^2$,
\be
F\;\sim\; \epsilon N + g_s^2 \epsilon N + \cdots\;.
\label{expansion}
\ee

When the scaling of the various parameters is appropriately fixed, the
bound state appears from  the outside to be indistinguishable from a BH. Since this collapsed-polymer model so immaculately replicates the  properties of a classical BH (and also those of a semiclassical BH \cite{emerge}),  one might wonder if there is still some property that allows one to distinguish the two descriptions. As shown in the main text,  this question can be answered affirmatively by comparing the QNM spectrum of the collapsed-polymer model with the conventional one for the BHs of  GR.

\subsection{Quasinormal modes}

Just like in the main text, we are limiting considerations to Schwarzschild BHs, even though rotating Kerr BHs are more realistic.

As is now well known (but see \cite{QNMBH,QNMBH2,Konoplya:2011qq} for reviews), a perturbed BH will settle down to its  equilibrium state  by ``ringing'' at characteristic complex frequencies which are determined  by only a handful of parameters. Since   a Schwarzschild BH has  only one characteristic scale, the frequencies are determined solely by the horizon radius $R_S$ or, equivalently, the surface gravity $\;\kappa=1/(2R_S)\;$. For both tensor  and scalar perturbations, the real parts are of order $\kappa$ for  all modes with low angular momentum
$\;\ell\sim 1\;$ (otherwise, the frequencies increase, roughly in proportion to $\ell$),
$\;\omega_R\equiv \text{Re}~\omega \sim \kappa\;$,  whereas the imaginary parts of the frequencies (or the inverses of the damping times) are, to a good approximation,  half-integer multiples of the surface gravity, $\;\omega_I\equiv\text{Im}~\omega \approx (m-1/2)\kappa\;$ with $\;m=1,2,3,\dots\;$. To be clear, this spectrum  has only been established rigorously in the large-$m$ or eikonal limit \cite{Motl}, although a WKB approximation attains roughly the same form at small $m$ \cite{wkb}, as do various numerical studies \cite{QNMBH,SchutzLRR}.

As shown in the main text, the appearance of a new scale in the polymer model is marked in a specific way in both the real and imaginary parts of the QNM spectrum.
It follows that  GW frequencies  could  provide a clear observational distinction between our model and classical BHs.

Two distinct notions of QNMs exist: the ``standard'' one that
is used, for example, in the description of quantum-optics and condensed-matter systems  (see, {\em e.g.}, \cite{Chin,Sett}) and there is also the BH notion of QNMs.  First, let us discuss the standard case.  Here, one is considering an open system that supports waves; for  instance, a dielectric or  an optical cavity, as either  provides a  partially reflecting outer surface. Such a system will lose energy to its environment, giving  rise to damped (complex-frequency) waves. To determine the QNM spectrum, one is instructed to impose
(1) totally reflecting boundary conditions at the center of the system and (2) the condition of purely outgoing waves in the external  environment and then, by
continuity, the same condition at the outer  surface of the system.
In effect, the exterior region is traced out of the problem. It is there only for conceptual reasons and plays  no essential role from a computational perspective.

The BH notion of  QNMs (see, {\em e.g.}, \cite{QNMBH,QNMBH2}) is different. For a BH spacetime, the problem can be set up like a scattering experiment, which is common in the high-energy literature ({\em e.g.}, \cite{niet}).  In this case,  one is considering  modes that initially came in from infinity and then were either reflected from or transmitted through the Schwarzschild potential barrier (at a radius of about $3/2~R_S$). The QNMs can be identified as  poles in the scattering amplitude, which is essentially a Fourier transform of the scattering potential. The boundary conditions are those of outgoing waves at spatial infinity and ingoing at the BH horizon.  Such a choice  of conditions suggests that it is now the interior which is, in effect, traced out, as it always is for an external observer in a BH spacetime. The setup for the BH QNMs is then, in some sense,  the mirror image of the standard description.

The simple model of Kokkotas and Schutz \cite{KSmodel} demonstrates how these two perspectives can both be accommodated. Those authors describe  the interior of some radiating system  as a finite string. This string is then coupled by a massless spring to a second,  semi-infinite string representing the exterior spacetime. The finite string will generally support two independent classes of modes;  one of which  is  coupled weakly to the exterior and another one,
coupled much more strongly.
Based on the discussion in \cite{KSmodel}, one might
expect that the modes of the former and latter classes are analogous to modes
from the standard and BH perspectives, respectively.
This  expectation has indeed been
verified by  studies on ultra-compact neutron stars and other (hypothetical) ultra-compact, relativistic stars ({\em e.g}, \cite{KSstar,AKK,inversecowling,Kokk1,Kokk2}~\footnote{Many more references can be found in the review articles
\cite{QNMBH,QNMBH2}.}). In these treatments, one  finds that the  $f$- and $p$-modes (meaning fundamental and pressure modes) are among those associated with the stellar fluid, whereas the so-called $w$-modes have more of resemblance  with   perturbations in the curvature of spacetime. The separation of the fluid modes from the spacetime modes is known as either the Cowling or inverse Cowling approximation \cite{cowling,inversecowling}.

One uses the Cowling approximation when  perturbations of the spacetime metric can be neglected. In this case, the strength of the coupling of the fluid  modes to the emitted GWs --- which in turn determines the amplitude of these emitted
waves --- can be  estimated by way of  the celebrated quadrupole formula, which treats the background spacetime as fixed and (essentially) flat \cite{ThorneRMP}.
In particular,
 $\;h\propto d^2Q/dt^2\;$, where
$h$ is the wave amplitude and $Q$ is the quadrupole moment of the  energy density.

The spectrum of QNMs of a BH-like object should be able to at least mimic the predominant modes  from the  spectrum of its  classical GR counterpart. The physical reason for this   is  that the associated ringdown process depends primarily  on the spacetime outside of the ultra-compact object, which must be indistinguishable from the exterior spacetime of a BH in GR. (The boundary conditions at the outer surface, which vary from model to model,  are also of relevance.)  However, if the interior of a BH-like object does contain some matter, then one would  expect, as discussed above, some additional (fluid) modes to be excited.  Classically, the fluid modes cannot couple to the spacetime modes in the presence of a horizon. However, quantum mechanically, fluid modes would be expected to  couple to the spacetime modes by way of ``quantum leakage'' and  then  propagate outside of the (would-be) horizon.

Just like for the modes of relativistic stars, the real part of the frequency  of a  QN fluid mode should be determined by the speed of  sound of the interior matter. This velocity  is necessarily less than but possibly saturating the speed of light $c$. For any BH-like object, the spatial scale of the interior is the Schwarzschild radius $R_S$; otherwise, the object is not sufficiently compact. It is then a generic result that the oscillatory frequency of a mode from this class is bounded from above, $\;\omega_R \leq c/R_S\;$. In addition, a time delay of order $1/\omega_R$  in the excitation of a mode can be  expected. This is because a waiting time of  at least one period is needed for this interior mode to affect the spectrum of QNMs outside of the BH and, therefore,  the spectrum of the emitted GWs.

As for  the damping time --- the inverse of the imaginary part of the
frequency  $\;\tau_{damp}=1/\omega_I\;$ ---  the situation is less conclusive. On general grounds, one might  expect the damping time of a fluid  mode  to be longer than those
of the spacetime modes \cite{KSstar}. To understand why, let us recall the quadrupole
formula, which says that the coupling to gravity of such modes is proportional to $\omega_R^2$. Then, since  $\;\omega_R<c/R_S\;$ is generically true,  their coupling must be weaker than it is for the relativistic spacetime modes. On the other hand, the intrinsic dissipation in the fluid could be strong, reducing the damping time.

In our model, the damping time of the matter modes is parametrically larger than $1/\omega_R$,
which can be attributed, in part,  to  the (normalized) intrinsic dissipation  being  very weak.
Because  of the  weak coupling of these modes to gravity,
the emission of GWs will take place
over an even longer  time scale and thus be a similarly weak source of dissipation. A longer damping time is consistent with  the expectations of Cardoso {\em et. al.} \cite{Cardoso}. The same authors also stressed  the importance of long-time observations in identifying deviations from GR.

\end{document}